\documentclass[12 pt, amsfonts, amssymb,color]{article}

\usepackage{amsmath,amssymb,amsfonts,latexsym,graphics,epsfig}
\usepackage{subcaption}
\usepackage{color}

\begin{document}
%<<<<<<<<<<< ennumeration of eqns section wise>>>>>>>>>>>>>>>>>>>

\renewcommand\theequation{\arabic{section}.\arabic{equation}}
\catcode`@=11 \@addtoreset{equation}{section}
%<<<<<<<<<<<<<<<<<<<<<<<<<<<<<<<<<>>>>>>>>>>>>>>>>>>>>>>>>>>>>>>>>>
\newtheorem{dfn}{Definition}[section]
\newtheorem{theo}{Theorem}[section]
\newtheorem{axiom2}{Example}[section]
\newtheorem{axiom3}{Lemma}[section]
\newtheorem{prop}{Proposition}[section]
\newtheorem{cor}{Corollary}[section]
\newcommand{\be}{\begin{equation}}
\newcommand{\ee}{\end{equation}}
\newcommand{\lmat}{\left(\begin{array}{cccccc}}
\newcommand{\rmat}{\end{array}\right)}
\newcommand{\tb}{\textbf}
\newcommand{\IG}{{\mathbb{G}}}
\newcommand{\p}{\partial}
\newcommand{\om}{{\Omega \cal G}}
\newcommand{\ID}{{\mathbb{D}}}
\newcommand{\pr}{\prime}

\title{Integrable modulation, curl forces and parametric Kapitza equation with trapping and escaping}

\author{Partha Guha$^{a}$\thanks{E-mail: {\tt partha.guha@ku.ac.ae}}, Sudip Garai$^{b}$\thanks{E-mail: {\tt sudip.dhwu@gmail.com}}\\
\\
$^{a}$Department of Mathematics \\
Khalifa University of Science and Technology \\ P.O. Box -127788, Abu Dhabi, UAE\\
\\
$^{b}$Department of Physics, \\
Diamond Harbour Women's University, \\
D. H. Road, Sarisha, West-Bengal 743368, India
}

\bigskip

\date{\today}

\maketitle

\smallskip

\begin{abstract}
{In this present communication the integrable modulation problem has been applied to study parametric extension of the
Kapitza rotating shaft problem, which is a  protypical example of curl force as formulated by
Berry and Shukla in [\emph{J. Phys. A} $\mathbf{45}$ 305201 (2012)]
associated with simple saddle potential. The integrable modulation problems yield parametric time dependent integrable systems. The Hamiltonian and first integrals of the linear and nonlinear parametric Kapitza equation (PKE) associated with simple and monkey saddle potentials have been given. The construction has been illustrated by choosing $ \omega(t)=a +b\cos t$ and that maps to Mathieu type equations, which yield Mathieu extension of PKE. We study the dynamics of these equations. The most interesting finding is the mixed mode of particle trapping and escaping via
the heteroclinic orbits depicted with the parametric Mathieu-Kapitza equation which are absent in the case of non parametric cases.}
 \end{abstract}


\smallskip


\smallskip

\paragraph{Keywords and Key phrases:}  First integrals; Curl force; Eisenhart-Duval lift; Kapitza equation;
Higher-order saddle potentials; Mathieu equation;  Heteroclinic orbits; Particle trapping and escaping;

\section{Introduction}

Berry and Shukla \cite{BS466,BS3} introduced Newtonian dynamics driven by forces $F({\bf r})$ depending only on position ${\bf r}$
and not on velocity ${\bf v}$ such that whose curl
is not zero, hence they are not derivable from a scalar potential. The curl flow preserves volume in the position-velocity
phase space $({\bf r}, {\bf v})$ where there are no attractors.
\noindent
Given a curl force (assuming unit mass for convenience)
$\ddot{\bf r} = F({\bf r})$, $\nabla \times F({\bf r}) \neq 0$, we obtain
$$
\nabla_{\bf r}{\bf v} + \nabla_{\bf v}\dot{\bf v} = \nabla_{\bf r}{\bf v} + \nabla_{\bf v}F({\bf r}) = 0,
\qquad {\bf v} = \dot{\bf r}.
$$

\smallskip

In early 90s  Moser and Veselov \cite{Veselov} considered the family of
integrable Hamiltonians $H(p,x;\Omega)$ depending on parameters $\Omega = (\omega_1,\omega_2, \cdots \omega_n)$,
which they made the parameters time dependent $\Omega = \Omega(t)$ with the time period $T$.
In general this destroys the integrability structure of the system, they considered a special case,
which they called integrable modulations, when the values of the first integrals $I_i(p(t),x(t);\Omega(t))$
of the modulated system
\be
\frac{dx}{dt} = \frac{\partial H}{\partial p}(p,x;\Omega(t)) \qquad \frac{dp}{dt} = -\frac{\partial H}{\partial x}(p,x;\Omega(t)),
\ee
are also $T$-periodic. This implies the time $T$-shift along the trajectories of the modulated
system gives an integrable symplectic map with the same integrals $I_1, \cdots , I_n$.

\smallskip

Let us illustrate with the harmonic oscillator. The linear harmonic oscillator has been a time honored favorite
and has  enhanced our understanding of several key areas of
mathematics and physics. Let $ H = \frac{1}{2}(p^2 + \omega^2 x^2)$ be the Hamiltonian of the harmonic oscillator,
then the corresponding Hamiltonian of the integrable modulated system is given by
\be\label{ham} H = \omega(t) \big( p^2 + x^2) \omega(t)H_0, \ee
where $H_0 = \frac{1}{2}(q^2 + p^2)$ is the Hamiltonian
for $\dot{q} = p$ and $\dot{p} = -q$.
This is integrable for any modulation for any $\omega (t)$.
By changing time $t \to \tau(t)$, a Sundman time, we write $d\tau = \omega(t) dt$.
Sundman time is a nonlocal transformation of time.
It is noteworthy
that the Hamiltonian (\ref{ham}) is not an integral of motion anymore, changes
periodically with time provided the modulation if $\omega(t)$ is periodic.

\smallskip

A few years before the Veselov's article,  Bartuccelli and
Gentile \cite{BG,BGW} independently  made a beautiful observation about this integrable modulation
problem. They presented a very elegant algorithm method to compute
the first integrals of the integrable modulated equations for oscillator and pendulum
problems. Later PG and his collaborator A. Ghose-Choudhury extended to other cases,
including the famous Mathieu equation \cite{CG14}. Integrable modulation technique has been also applied
to obtain the first integrals of the Emden-Fowler equation \cite{GC}.

\smallskip

In general such kind of modulation does not exist for generic integrable
systems, for example, this does not exist for the Neumann system or
for the closely related Jacobi system of geodesics on ellipsoids. But Veselov \cite{Veselov}
showed that this modulation exits for Euler's rigid body dynamics and also to its
$N$--dimensional generalization. In fact, Veselov demonstrated in \cite{Veselov} that
the modulated Euler system can be written in the Lax form
\be
\dot{L} = [L,P], \quad \hbox{ with } \,\,\, L = M + \lambda J_{0}^{2}, \,\,\,\, P = \Omega + \lambda J,
\ee
where $M \in so(N)^{\ast}$, $\Omega \in so(N)$ and $J$ is a symmetric matrix
$J = (J_{0}^{2} + f(t){\Bbb I})^{1/2}$, where ${\Bbb I}$ is the identity matrix. Here $J$ is defined via
an arbitrary scalar function $f(t)$ and a constant symmetric matrix $J_0$.
It is clear that the integrals of the non-perturbed system are preserved by the modulated system.

\bigskip

Our work lies at the crossroad of two main ideas, in one hand it is connected to integrable modulation
as proposed by Moser and Veselov and on the other hand it is related to Berry-Shukla's work \cite{BS466,BS3} on
curl forces, which leads to the famous Kapitza equation, where the underlying potential is the saddle potential \cite{Guha1}.
The curl force plays an important role in optical trapping and PT symmetric systems \cite{Guha2}.
Recently dynamics of the curl force associated to the higher-order saddles have been studied,
using the pair of higher-order saddle surfaces and rotated saddle surfaces a generalized rotating shaft equation
is constructed \cite{GaraiGuha}

As we have seen from the paper of Veselov that the integrable modulations do not exist for most of the systems.
In this paper we show that such modulation exists for (higher ) saddle potentials, which in turn, related to Kapitza-Merkin equation \cite{Kapitsa,Merkin,Merkin1}. We use the method described in \cite{GaraiGuha} to construct parametric extension of the (generalized) Kapitza-Merkin equations, which satisfy integrable modulation. Using a particular ( or periodic) choice of $\omega(t)$ we derive a Mathieu type extension of the Kapitza equation which we coin as Mathieu-Kapitza-Merkin equation. We also show that the Eisenhart-Duval lift is one of the best way to describe a geometric description of the integrable modulated equations.
The notable contribution of this present work is to observe the mixed phase space of trap and escape for
the particles via the parametric curl forces. In the absence of the parametric influence on the nonlinear
curl forces the trapping is most obvious and the phase space trajectories are self-retracing and
self crossing on most of the times, whereas, in the parametric cases the phase space plots depicted homoclinic,
heteroclinic and mixed mode of cycles.

\bigskip

This paper is {\it organized} as follows. We describe integrable modulation, parametric differential equations and corresponding
first integrals in Section 2. We also give a geometric interpretation of modulated systems in terms of Eisenhart-Duval
lift. The parametric Kapitza equation is explored in Section 3 as an example of integrable modulated system.
Section 4 is dedicated to Mathieu extension of the Kapitza equation for a periodic value of $\omega (t)$. We also study numerically
the dynamics of these equations and demonstrate the trapping and escaping phenomena in this section.

\section{Preliminaries: Integrable modulated equation and first integrals}

We first briefly outline the method of construction of first integrals of a class of integrable modulated equations
which we will study later.\\
Bartuccelli and Gentile made an interesting
observation in \cite{BG} regarding the parametric modulational
 equation of a linear harmonic oscillator
\be\label{1.1} \ddot{x}+\omega^2 x=0.\ee  As it is well known the
solution of the harmonic oscillator is $ x(t)=A\cos(\omega t+\phi)$ , where
$A$ and $\phi$ are arbitrary constants representing the amplitude
and phase respectively. Moreover the energy integral
\be\label{1.1b}E=\frac{1}{2}\dot{x}^2+\frac{1}{2}\omega^2 x^2\ee
is a constant of motion. They observed that if one
assumes that $\omega$ instead of being a constant is any arbitrary
function of the independent variable $t$ so that the solution of
the following equation:
\be\label{1.3}
\frac{d}{dt}\left(\frac{\dot{x}}{\omega(t)}\right)+\omega(t) x=0,\ee
has similar structure to usual oscillator equation in the sense
that  it is given by \be\label{1.3a} x(t)=A \cos\left(\int^t \omega(t^\prime)
dt^\prime +\phi\right).\ee In parallel to the energy integral
when $\omega$ is explicitly dependent on time a first
integral of (\ref{1.3}) is given by
\be E(x,\dot{x},t)=\frac{1}{2}\left[\left(\frac{\dot{x}}{\omega(t)}\right)^2+x^2\right].\ee
 It is obvious that (\ref{1.3}) is not expressible in Hamiltonian equation
 of the form  (1.1).
 %$\ddot{x}+\omega^2(t)x=0$.
  Nevertheless the fact that the
 solution (\ref{1.3a}) clearly reduces to that of the usual
 harmonic oscillator when $\omega$ is a constant. In fact the following generalization of (\ref{1.3}),
\be\label{1.4}
\frac{d}{dt}\left(\frac{\dot{x}}{\omega(t)}\right)+\omega(t)
F(x)=0,\ee is also possible where $F(x)$ is some nonlinear $C^1$
function of $x$. The canonical equations of motion of (\ref{1.4})
is \be \label{1.4a} \dot{x} = \frac{\partial H}{\partial p} = \omega(t) p, \qquad \dot{p}
= -\frac{ \partial H}{\partial x} = - \omega(t)\frac{\partial U}{\partial x}, \ee where
the Hamiltonian is given by $H = \frac{1}{2}\omega(t)\big(p^2 + U(x)\big)$. Thus
$U(x)$ is the primitive of $F(x)$. It is clear that the Hamiltonian is no longer a
conserved quantity.

\smallskip

The structural similarity of the
solution and first integral of (\ref{1.1}) and (\ref{1.3})
constitutes the essential feature of Bartucelli and Gentile's
observations. Equation (\ref{1.4}) has a first integral  given by
\be\label{1.5} I(x(t),\dot{x}(t),t) =
\frac{1}{2}\left(\frac{\dot{x}}{\omega(t)}\right)^2 + U(x(t)). \ee
Multiplying (\ref{1.5}) by $\omega^2(t)$ and doing some
rearrangement one obtains \be\label{1.6} \int \frac{dx}{\sqrt{E -
U(x)}} = \pm \sqrt{2}\int dt\, \omega (t). \ee When we restrict
ourselves to one sign then the right hand side of (\ref{1.6}) involves
effectively a re-parametrization of the independent time variable.
In general however, since $\omega(t)$ can change sign, it is still
possible for the ratio $\dot{x}(t)/\omega(t)$ to be well defined
since $x(t) = A \cos\big(\int \omega(t) dt\big)$. Hence
(\ref{1.3a}) and (\ref{1.6}) may be regarded as the natural
extension of  $\omega=$constant case to a time-dependent
$\omega(t)$.

\bigskip

{\bf Connection to adaptive frequency oscillator :}\,
It is noteworthy to mention that the integrable modulation problem finds an application to
design a learning mechanism  for oscillators, which  adapts the oscillator frequency to the frequency of any periodic input signal.
In an interesting paper Ijspeert and his coworkers \cite{Auke} showed that the adaptation  mechanism causes an oscillator’s frequency to converge to the frequency of any periodic input signal, for phase and Hopf  oscillators.
The corresponding equations for the adaptive  Hopf oscillator are given by
\be
\dot{x} = (\mu - x^2 - y^2)x - \omega y + \epsilon F(t), \,\,\,\, \dot{y} = (\mu - x^2 - y^2)y + \omega x, \,\,\,\,
\dot{\omega} = - \epsilon F(t)\frac{y}{\sqrt{x^2 + y^2}},
\ee
where the adaptive frequency oscillator rotates around limit cycle in the xy-plane. We consider the situation
for $x^2 + y^2 = \mu$, then from the first two equations we obtain
\be
\frac{\dot{x}}{\omega} + \omega x = -\frac{\epsilon F(t)}{\omega^2} + \frac{\epsilon \dot{F}(t)}{\omega}.
\ee

\subsection{Eisenhart-Duval lift and geometric description }

The Eisenhart-Duval lift \cite{Eisen,Duval1,Duval2,Cariglia}
provides a nice geometric description of a differential equation with $d$-degrees of freedom
$q_i, i = 1, \cdots , d$ and the potential energy $U(t,q)$ in terms of geodesics of he Lorentzian metric on a
$(d+2)$-dimensional space-time
\be
g_{\mu \nu}(x)dx^{\mu}dx^{\nu} = dq_idq_i - dtdv - 2U(t,q)dt^2,
\ee
where $x^{\mu} = (t,v,q_i)$.  Let us recall the Hamiltonian of the time-dependent oscillator
\be H = \frac{p_{q}^{2}}{2m(t)} + \frac{1}{2}m(t)\omega^2(t)q^2.
\ee
write the Hamiltonian ${\cal H}$ for the extended phase space. Let $H -p_t$, then
\be
{\cal H} = 2p_tp_v + \frac{p_{q}^{2}}{m} + m\omega^2q^2p_{v}^{2} = g^{\mu \nu}p_{\mu}p_{\nu},
\ee
thus the corresponding Eisenhart-Duval metric is given by
\be\label{metric}
ds^2 = g_{\mu \nu} dx^{\mu}dx^{\nu} = 2dtdv + mdq^2 - m\omega^2q^2dt^2.
\ee

Let us set $\tilde{q} = \frac{q}{y(t)}$.  In this new coordinate metric (\ref{metric}) can be expressed as
\be
ds^2 = 2dt(dv + \lambda) + \frac{my^2}{\omega^2}\tilde{q}^{2},
\ee
where
\be
\lambda = my\dot{y}\tilde{q}d\tilde{q} + \frac{1}{2}m\tilde{q}^{2}(\dot{y}^{2} - y^2\omega^2)dt
\ee
is a one-form.
This follows from
\be dq^2 = \big(d{\tilde{q}}y + \tilde{q}\dot{y}dt)^2 = y^2 {d\tilde{q}}^2 + 2y\dot{y}\tilde{q} d\tilde{q}dt
+ {\tilde{q}}^2\dot{y}^{2} dt^2.
\ee

\begin{prop}
The closedness of $\lambda$ implies
\be
	\frac{d}{dt}(m\dot{y}) + m\omega^2(t)y = 0.
\ee
\end{prop}

{\bf Proof :}\,  If $\lambda$ is closed then $d\lambda = 0$, latter implies
\begin{gather}
d\lambda = \big(\frac{d}{dt}(m\dot{y}) + m\dot{y}^{2} \big)\tilde{q} dt \wedge d\tilde{q}
+ m(t)\dot{y}^{2}\tilde{q} d\tilde{q} \wedge dt - m(t){y}^{2}\omega^2(t)\tilde{q} d\tilde{q} \wedge dt \nonumber \\
= - \big(\frac{d}{dt}(m\dot{y}) +  m(t){y}^{2}\omega^2(t)  \big)\tilde{q}d\tilde{q} \wedge dt.
\end{gather}
$\Box$

\smallskip

{\bf Remark :}\, When we set $m(t) = 1/ \omega(t)$, we recover our integrable modulated oscillator equation
which yields the usual integral of motion, but this is not available for the generic cases.

\smallskip

One can generalize this to arbitrary function $F = F(y)$. Let us set
\be \tilde{q} = \frac{q}{F(y)}. \ee
This transforms the metric to
\be
ds^2 = 2dt(dv + \Lambda) + mF^2 dQ^2,
\ee
where
$$ \Lambda = mFF^{\prime}\dot{y}QdQ + \frac{1}{2}mQ^2\big({F^{\prime}}^2\dot{y}^{2} - F^2\omega^2)dt^2. $$
If we demand $\Lambda$ to be closed, then  $d\Lambda = 0$ yields
\be
\frac{d}{dt}(mF^{prime}\dot{y}) + m\omega^2 F = 0, \quad \hbox{ or } \quad \frac{d}{dt}(m\dot{F}) +  m\omega^2 F = 0.
\ee

\section{Parametric Kapitza equation and integrable modulation}

The linearized dynamics of a rotating shaft formulated by Kapitsa \cite{Kapitsa} is given by
\be \label{Kapitsa}
\ddot{x} + ay + bx = 0, \qquad \ddot{y} - ax + by = 0.
\ee
The corresponding characteristic equation shows that the addition of a non-zero nonconservative
curl force ( i.e. $ a \neq 0$) to a stable system with a stable potential energy makes it unstable.
This is connected to Merkin's result \cite{Merkin,Merkin1}, which states that the introduction of nonconservative linear forces
into a system with a stable potential and with equal frequencies destroys the stability regardless of the form
of nonlinear terms. It is worth mentioning that the positional force, i.e.
the terms $ay$ and $-ax$ are proportional to $\omega^2$ ,
where $\omega$ is the rotation rate of the shaft. If we express (\ref{Kapitsa}) in a matrix form,
the potential part, $B$, is a diagonal matrix with equal
eigenvalues, $b$, and non-conservative part, $A$ , is an skew-
symmetric matrix. It is easy to see from the corresponding characteristic equation, $\hbox{ det }\big((\lambda^2 + b){\Bbb I} + A \big) = 0$, which implies $\lambda^2 + b$ is imaginary, thus we say that $\lambda$ is unstable.

\bigskip

The equation (\ref{Kapitsa})  associated to a  saddle potential also can also be derived via Euler-Lagrange method.
It is straight forward to check that the  Lagrangian \be L = \frac{1}{2}(\dot{x}^{2} -
\dot{y}^{2}) - \frac{1}{2}a(x^2 - y^2) - bxy \ee
yields (\ref{Kapitsa}), and the corresponding Hamiltonian is given by
\be H = \frac{1}{2}(p_{x}^{2} - (p_{y}^{2}) + \frac{1}{2}a(x^2 - y^2) + bxy. \ee

It is noteworthy to say that the potential $(x^2 - y^2)$ is a simple saddle,
this  function  is  also  known  as  a hyperbolic  paraboloid.
The  function $f(x,y) = 2xy$ is  a  rotated  version  of  the same  surface.
Here the kinetic energy part of the Lagrangian is the anisotropic one.

\subsection{Parametric Kapitza equation}

Let us consider $\omega = \omega(t)$, then the parametric Kapitza equation is given by
\be
\frac{d}{dt}\big(\frac{\dot{x}}{\omega(t)}\big) + \omega(t)\,x + \omega(t)\,y = 0, \,\,\,\,\,\,
\frac{d}{dt}\big(\frac{\dot{y}}{\omega(t)}\big) + \omega(t)\,y - \omega(t)\,x = 0.
\ee
This pair of equation boils down to standard Kapitza equation of equal coefficients ($a=b = \omega$)
for constant $\omega$. This system of second-order differential equations can be derivable from a Lagrangian.

\smallskip

\begin{prop}
The Euler-Lagrange equation of the parametric Lagrangian
\be\label{paraLag}
	L = \omega(t) \Big(\frac{1}{2}\big(\frac{\dot{x}}{\omega(t)}\big)^2 - \frac{1}{2}\big(\frac{\dot{y}}{\omega(t)})^2
- \frac{1}{2}(x^2 - y^2 \big) - xy \Big)\ee
yields the parametric Kapitza equation.
\end{prop}

The equation can be manufactured from the Euler-Lagrange equation using time dependent simple
saddle potential $g_1 (x, y) = \frac{1}{2}\omega(t)(x^2 - y^2)$ and the rotated version of the same surface,
$g_{1}^{r}(x, y) = \omega(t)xy$

\smallskip

\begin{prop}
The first integral of the parametric Kapitza equation is given by
	\be\label{firstint1}
	I_1 = \frac{1}{2}\big((\frac{\dot{x}}{\omega(t)})^2 - (\frac{\dot{y}}{\omega(t)})^2 \big)
	+ \frac{1}{2}(x^2 - y^2) + xy.
	\ee
	\be\label{firstint2}
	I_2 = \frac{\dot{x}}{\omega(t)}\frac{\dot{y}}{\omega(t)} - \frac{1}{2}(x^2 - y^2) + xy.
	\ee

\end{prop}

{\it{Proof }} : It is straight forward to check
\begin{gather}
\dot{I} = \frac{\dot{x}}{\omega(t)}\frac{d}{dt}\big(\frac{\dot{x}}{\omega(t)}\big) -
\frac{\dot{y}}{\omega(t)}\frac{d}{dt}\big(\frac{\dot{x}}{\omega(t)}\big)
+ x\dot{x} - y\dot{y} + \dot{x}y + x\dot{y} \nonumber \\
= \frac{\dot{x}}{\omega(t)}\big( - \omega(t) x - \omega(t) y \big) - \frac{\dot{y}}{\omega(t)} \big(
- \omega(t) y + \omega(t) x) + x\dot{x} - y\dot{y} + \dot{x}y + x\dot{y} = 0
\end{gather}

Similarly we can prove the second integral of motion $I_2$ using direct computation.

$\Box$

\bigskip

The immediate generalization of simple saddle is known as monkey saddle.
This saddle is so-named because it could be used by a
monkey; it has places for two legs and a tail.
We can generalize rotating shaft equation formulated by Kapitza using higher-order saddles.
We will tacitly use the construction given in \cite{GaraiGuha}, here we give a parametric generalization of
our previous result.

\bigskip

The generalized rotating shaft equation
associated to degree 3 is given by
$g_2 (x, y) = (x^3 - \frac{1}{3}xy^2)$ and the corresponding rotated version
$g_{1}^{r}(x, y) = (x^2y - \frac{1}{3}y^3)$.
\be
\ddot{x} + \omega^2(x^2 - y^2) + 2\omega^2 xy = 0, \quad
\ddot{x} - \omega^2(x^2 - y^2) + 2\omega^2 xy = 0.
\ee

\smallskip

Now we demand $\omega = \omega(t)$, hence the potential is the time dependent one. The parametrized equation is given by
\be\label{paraKapitza}
\frac{d}{dt}\big(\frac{\dot{x}}{\omega(t)}\big) + \omega(t)\,(x^2 - y^2) + 2\omega(t) xy = 0, \,\,\,\,\,
\frac{d}{dt}\big(\frac{\dot{y}}{\omega(t)}\big) - \omega(t)\,(x^2 - y^2) + 2\omega(t) xy = 0.
\ee

\begin{prop}
The first integrals of the generalized parametric Kapitza equation are given by
	\be\label{Firstint1}
I_1 = \frac{1}{2}\big((\frac{\dot{x}}{\omega(t)})^2 - (\frac{\dot{y}}{\omega(t)})^2 \big)
	+ (\frac{1}{3}x^3 - xy^2) + ( x^2y - \frac{1}{3} y^3).
	\ee
\be\label{Firstint2}
	I_2 = \frac{\dot{x}}{\omega(t)}\frac{\dot{y}}{\omega(t)} +
	( x^2y - \frac{1}{3} y^3)
	- (\frac{1}{3}x^3 - xy^2).
	\ee
\end{prop}

{\it Proof } \, This can be proved by a direct calculation. $\Box$

\smallskip

This idea can be generalized to other higher saddle potentials also.

\subsection{Hamiltonian form and integrable modulation}

We compute momenta using Legendre transformation from the parametric Lagrangian (\ref{paraLag})
\be\label{mom}
p_x = \frac{\dot{x}}{\omega(t)}, \qquad p_y = -\frac{\dot{y}}{\omega(t)}.
\ee
Plugging into $I_1$ and $I_2$ we can express both the integrals of motion of the Kapitza-Merkin equation,
given in proposition 3.2, in terms of phase space coordinates
\be
\tilde{I}_1 \equiv H = \frac{1}{2}(p_{x}^{2} - p_{y}^{2}) + \frac{1}{2}(x^2 - y^2) + xy, \quad
\tilde{I}_2 = -p_xp_y - \frac{1}{2}(x^2 - y^2) + xy.
\ee

\bigskip

We now check the properties of the Hamiltonian and integrals of motion of the parametric system as described by Veselov.

\smallskip

\begin{prop}
The Hamiltonian equations of the  parametric Kapitza equation is given
	\be {\cal H} = \omega(t) \big(\frac{1}{2}(p_{x}^{2} - p_{y}^{2}) + \frac{1}{2}(x^2 - y^2) + xy)
	= \omega(t)H, \ee
where $H$ is the Hamiltonian of the Kapitza equation. The Hamiltonian equation yields
	\be \dot{x} = \frac{\partial {\cal H}}{\partial p_x} =  \omega(t)p_x, \quad \dot{p_x}
	= - \frac{\partial {\cal H}}{\partial x} = -\omega(t)\big(x + y \big),
	\ee
	\be \dot{y} = \frac{\partial {\cal H}}{\partial p_y} = -\omega(t)p_y, \quad \dot{p_y}
	=  - \frac{\partial {\cal H}}{\partial y}
	= -\omega(t)\big(y -x \big). \ee
	and this set of equations satify structure of the integrable modulation Hamiltonian equations.
\end{prop}
It can be easily shown that this set of Hamiltonian equation yields the parametric extension of the
Kapitza equation and fulfills the requirement of integrable modulation, i.e.,
${\cal H} = \omega(t)H$, where $H$ is the Hamiltonian
of the Kapitza equation without coefficients
$H = \frac{1}{2}(p_{x}^{2} - p_{y}^{2}) + \frac{1}{2}(x^2 - y^2) + xy$.

\bigskip

The first integral of motion  $\tilde{I}_1 \equiv H$ is no longer Hamiltonian, it has to be modulated by $\omega(t)$.
Let us express the second integral of motion  $\tilde{I}_2$ in modulated form as
${\cal H}_2 = \omega(t) \tilde{I}_2$, this
commutes with the Hamiltonian ${\cal H}$
$\{ {\cal H}_1, {\cal H}_2 \} = 0$, where $\{., .\}$ is standard Poisson bracket

\bigskip

\begin{prop}
	The parametric generalized Kapitza equation associated to the monkey saddle potential
	can also be expressed in terms of following Hamiltonian
\be
	{\cal H}_{MS} = \omega(t) \big(\frac{1}{2}(p_{x}^{2} - p_{y}^{2})
	+ (\frac{1}{3}x^3 - xy^2) + ( x^2y - \frac{1}{3} y^3) \equiv \omega(t) I_1,
\ee	
	where $I_1$ is the first integral or Hamiltonian of the generalized Kapitza equation.
\end{prop}
	
Also the second integral of motion can be expressed as
\be
{\cal H}_2  = \omega(t) \Big( p_xp_y  +
	( x^2y - \frac{1}{3} y^3)
	- (\frac{1}{3}x^3 - xy^2)\Big).
\ee

\section{The Mathieu equation}

The standard Mathieu equation \be\label{M.1} \ddot{x}+
(a+b\cos\Omega t)x = 0, \ee has $ a $ and $b \in {\Bbb R}$.  We
assume that $a>b>0$. Now, Floquet's theorem asserts that the
periodic solutions of the Mathieu equation can be expressed in
form $x(t) = X_1(t) = e^{i\mu t} p(t)$ or $x(t) = X_2 =  e^{-i\mu
t} {p}(-t)$. The constant $\mu$ is called the characteristic
exponent and $p(\pm t)$ is a $\pi$-periodic function of $t$. If
$\mu$ is an integer, then $X_1(t)$ and $X_2(t)$ are linearly dependent
solutions. Moreover, $x(t + k\pi) = e^{i\mu k \pi}x(t)$ or $x(t +
k\pi) = e^{-i\mu k \pi}x(t)$ stand for periodicity of the solutions $X_1(t)$ and
$X_2(t)$ respectively. On the other hand, the general solution of
the Mathieu equation for non integer values of $\mu$ is given by
\be x(t) = C_1  e^{i\mu t} p(t) + C_2e^{-i\mu t} {p}(-t), \ee where $C_1$ and $C_2$ are arbitrary constants.
It is also known that when $\mu$ is complex, i.e., $\mu = k + il$, then  for $ l\neq 0$
one obtains an unbounded solution to the Mathieu equation, while
for purely imaginary  values of $\mu$ one obtains real, bounded,
oscillatory solutions for $x(t)$ (see for example,
\cite{Arscott}).

%==============================
\begin{figure}[h]
\begin{subfigure}{.5\textwidth}
  \centering
  \includegraphics[width=5.9 cm]{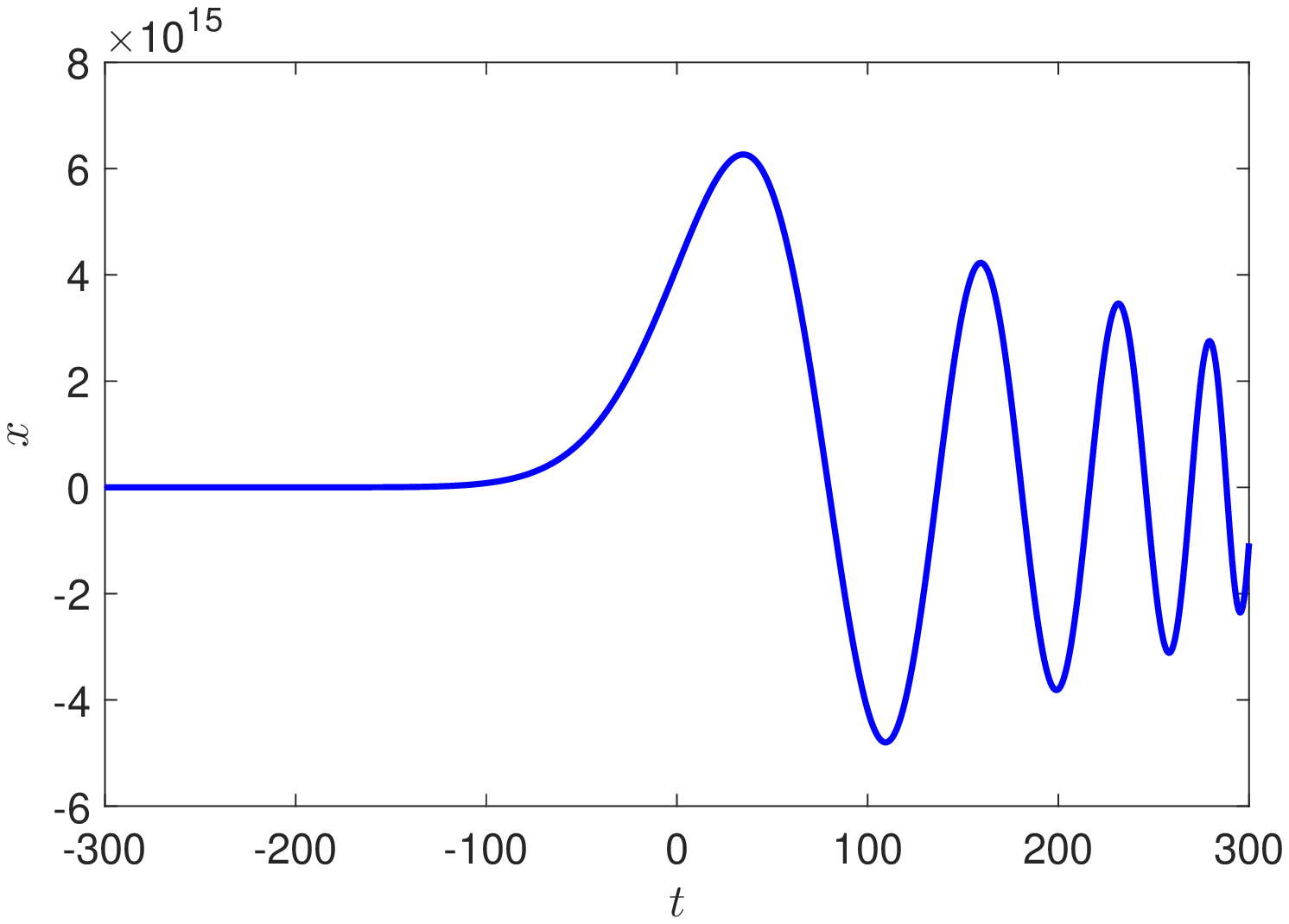}
  \caption{Plot of $x$ vs $t$.}
  \label{fig:sub-third3}
\end{subfigure}
\begin{subfigure}{.5\textwidth}
  \centering
  % include fourth image
  \includegraphics[width=5.9 cm]{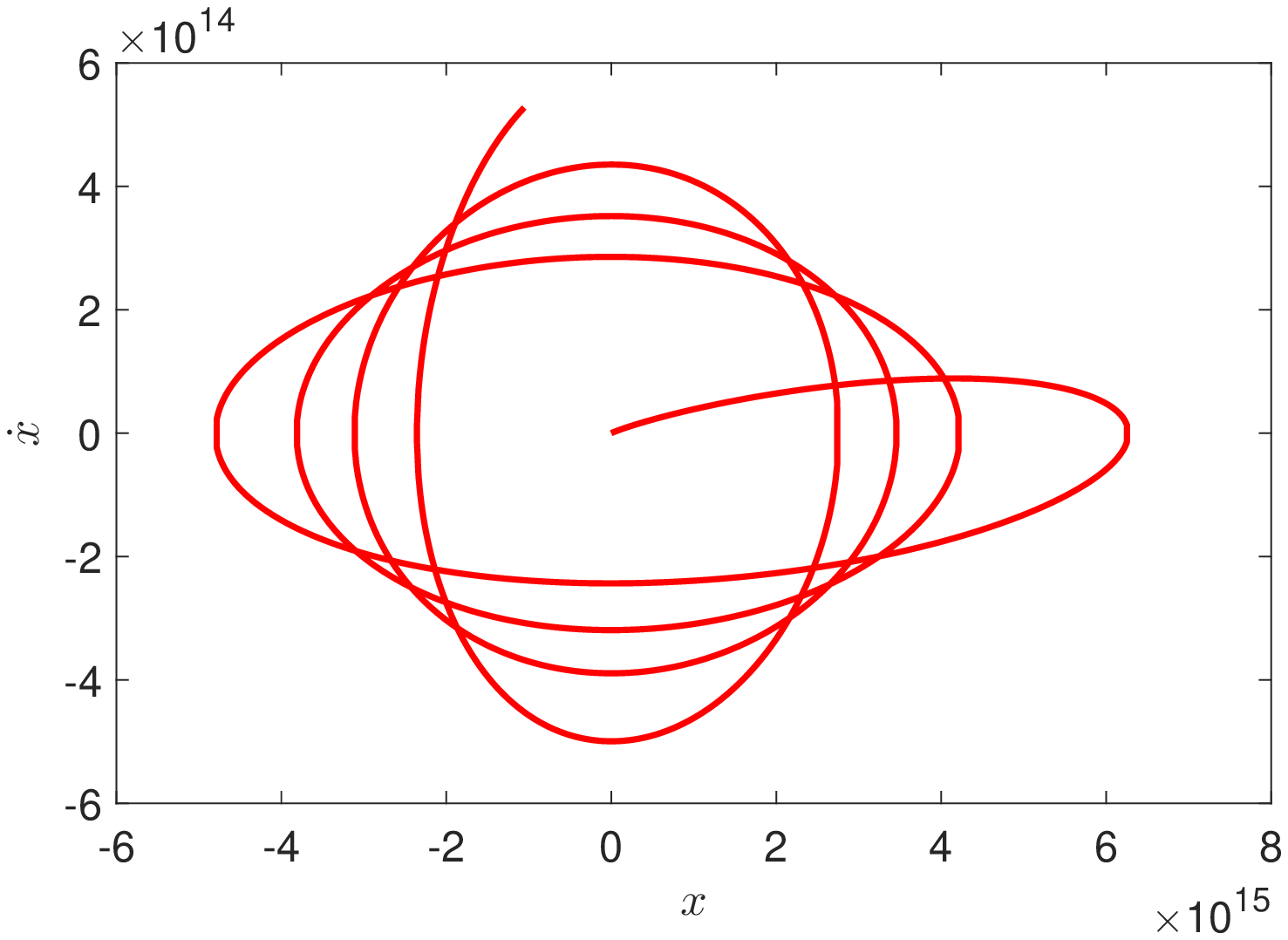}
  \caption{Plot of $\dot{x}$ vs $x$.}
  \label{fig:sub-fourth3}
\end{subfigure}
\caption{Trajectories and phase space diagram solved from Eq.(\ref{a3}) with $q =0.1$, $a= 0.01=b$,$\alpha_1=0.01=\alpha_2$, $\Omega=0.01$. The initial conditions are $x(0)=0$, $\dot{x}(0)=0.1$.}
\label{f1}
\end{figure}
%=================================================================
%

In order to study the integrable modulated Mathieu equation (\ref{M.1}) which leads to
a special kind of time-dependent damping, at first we fix
\be \omega^2(t)=a+b\cos\Omega t. \ee
Since
$$\frac{d}{dt}\left(\frac{\dot{x}}{\omega(t)}\right)+\omega(t)x = \ddot{x}
- \frac{\dot{\omega}(t)}{\omega(t)}\dot{x} + \omega^2(t) = 0,
$$
hence the modulated Mathieu equation is given by

\be\label{a3}\ddot{x}+ \frac{\Omega
b\sin\Omega t}{2(a+b\cos\Omega t)}\dot{x}+ a+b\cos (\Omega t) x=0.\ee and set
$\omega^2(t)=a+b\cos (\Omega t)$.
This has a distinct
advantage for us, since it is already mentioned in (\ref{1.3}) that this admits the
following first integral, \textit{viz}
\be\label{a5}I(x,\dot{x},t)=\frac{1}{2}\left[\left(\frac{\dot{x}^{2}}{ a+b\cos (\Omega t) }\right)+x^2\right].\ee
Thus on the level surface $I=C_1$ one finds that the general
solution  of the integrable modulated Mathieu equation, which is given by
\be \int\frac{dx}{\sqrt{2C_1-x^2}}=\pm\int \sqrt{a+b\cos\Omega t}\;dt +
C_2,\ee
with $C_1$ and $C_2$ being  arbitrary constants. Explicit
evaluation of the integral on the left yields
\be x(t)=\sqrt{2C_1}\sin\left(\pm\int \sqrt{a+b\cos\Omega t}\;dt +
C_2\right).\ee
Note that the integral on the right hand side may be
expressed in terms of an elliptic integral of the second kind.

In figure (\ref{f1}), the trajectories and the phase space diagrams are depicted from the modulated Mathiew equation viz. Eq.(\ref{a3}). One can see that the periodic oscillation of the system i.e. the $x$ vs $t$ graph is somewhat damping in nature. This solution is analogous to the Airy function viz. $Ai(-t)$ as discussed by Berry\cite{bjpam}. In the limit, $a,b,\Omega \rightarrow 0$, Eq.(\ref{a3}) represents analogous $Ai$ function solutions.  On the other hand in the phase space diagram the periodic behavior is approached with many periods along with different amplitudes while the time part is changing from some negative finite value to positive finite value. The nature of this phase space is somewhat analogous to the functional plot of $Ai(-t)$ vs $Ai'(-t)$. This might be the direct consequence of the Eq.(\ref{a3}) under the previously said parametric range. This phase space diagram ensures the trapping of the charged particles within the domain of the periodic boundaries. These lines are not totally the heteroclinic orbits where the system starts from one saddle point and finishes with another saddle points without any kind of periodicity, instead, they are close to limit cycles.

\subsection{Damped extension of the nonlinear Mathieu equation}

In this section we extend the result of the previous section to a
damped counter part of the nonlinear version of the Mathieu
equation considered by Abraham and Chatterjee\cite{AC}. A weakly
nonlinear version of the Mathieu equation is applicable among
other things, also to Paul trap mass spectrometers which use static
direct current (DC) and radio frequency (RF) oscillating electric
fields to trap ions.

%==============================
\begin{figure}[h]
\begin{subfigure}{.5\textwidth}
  \centering
  \includegraphics[width=5.9 cm]{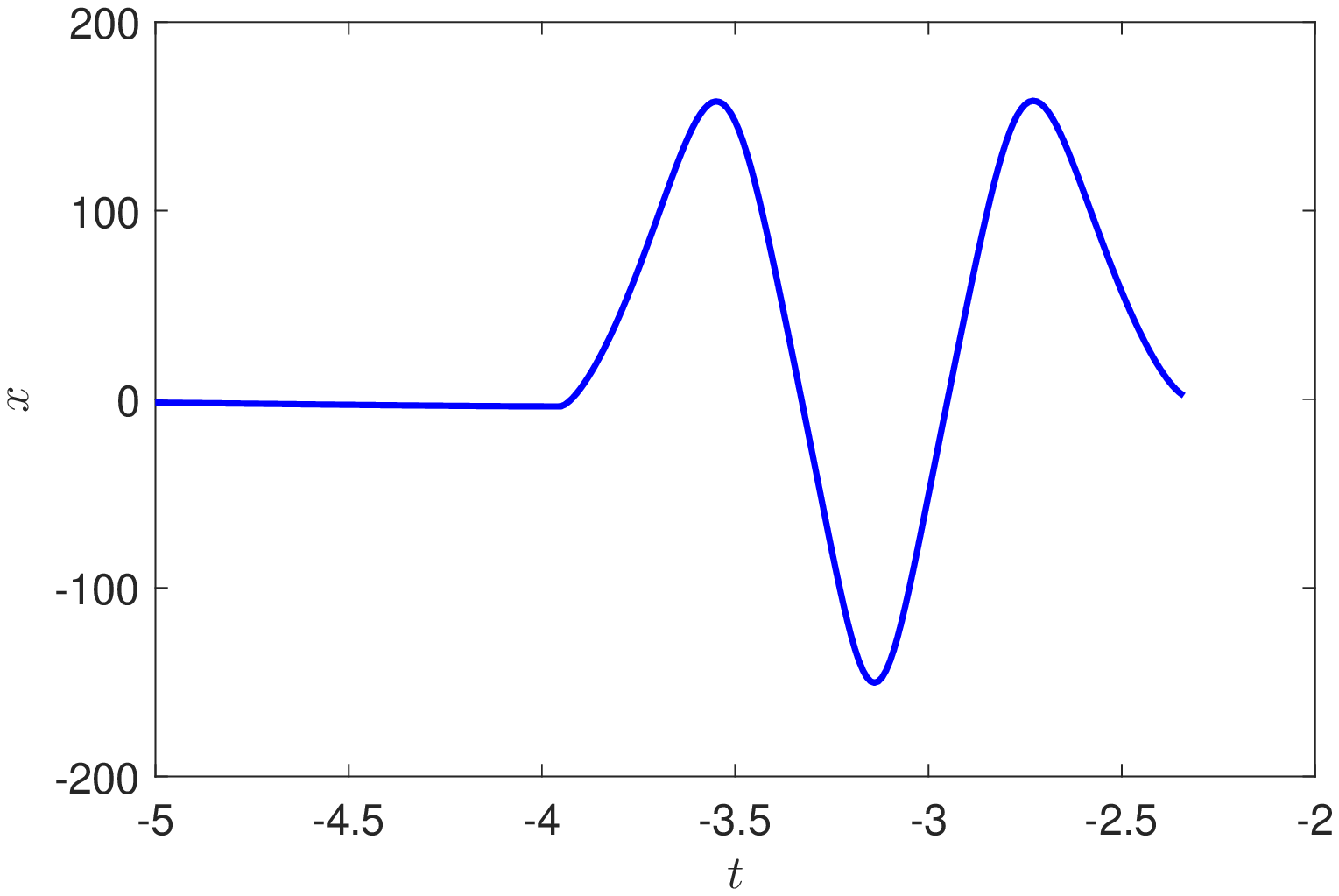}
  \caption{Plot of $x$ vs $t$.}
  \label{fig:sub-third3}
\end{subfigure}
\begin{subfigure}{.5\textwidth}
  \centering
  % include fourth image
  \includegraphics[width=5.9 cm]{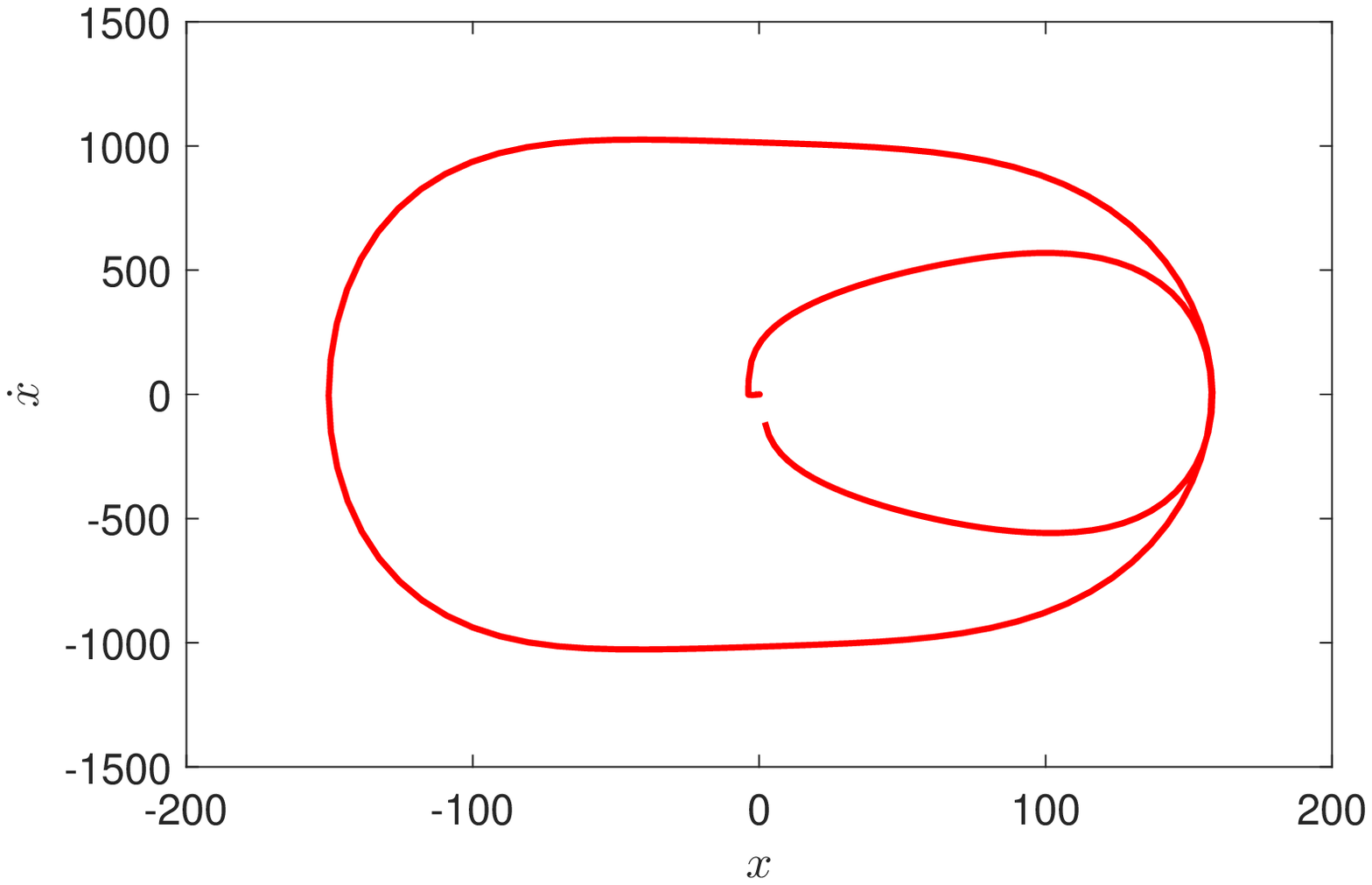}
  \caption{Plot of $\dot{x}$ vs $x$.}
  \label{fig:sub-fourth3}
\end{subfigure}
\caption{Trajectories and phase space diagram solved from Eq.(\ref{nm2}) with $q =0.1$, $a= 0.01$,$\alpha_1=0.1=\alpha_2$. The initial conditions are $x(0)=0$, $\dot{x}(0)=0.1$.}
\label{f2}
\end{figure}
%=================================================================
%

Consider a nonlinear Mathieu equation of the form \be\label{nm1}
\ddot{x} + (a + 2q \cos 2t)(x + \alpha_1x^2 + \alpha_2x^3) = 0,
\ee where the parameters $\alpha_1$ and $\alpha_2$ are small. We
 study this nonlinear Mathieu equation in presence of time
dependent damping, given by \be\label{nm2}  \ddot{x} +
\big(\frac{2q \sin\,2t}{a + 2q\cos\,2t} \big)\dot{x} + (a + 2q
\cos 2t)^2(x + \alpha_1x^2 + \alpha_2x^3) = 0. \ee

 In figure (\ref{f2}), we plotted the trajectory and the phase space for the Eq.(\ref{nm2}). The space part shows a pulse propagation. This solution is also analogous to the Airy function viz. $Ai(-t)$ given by the Eq.(\ref{nm2}) in the limit $q, a,\alpha_1,\alpha_2 \rightarrow 0$.\cite{bjpam} The phase space shows a periodic nature with double periods. The nature can be treated as the homoclinic as the system starts from one saddle point and almost comes back to that point again with the time running from some negative finite value to positive finite value. In the mean time it traces double periodic nature. One thing to notice here that the periods have different amplitudes and the charged particles can almost be trapped inside those two periodic boundaries.

Let us set
$\omega(t) = (a + 2q\cos\,2t)$, so that we may identify
(\ref{nm2}) with (\ref{1.4}) for $F(x) = x + \alpha_1x^2 +
\alpha_2x^3$, the corresponding potential function is given by
\be U(x) =  \frac{1}{2}x^2 +
\frac{\alpha_1}{3} x^3 + \frac{\alpha_2}{4}x^4). \ee

\smallskip

It is clear from the knowledge of earlier sections that the system (\ref{nm2}) has a first integral given by
\be\label{x} I =
\frac{1}{2} \Big[ (\frac{\dot{x}}{\omega(t)})^2 + (x^2 +
\frac{2\alpha_1}{3} x^3 + \frac{\alpha_2}{2}x^4 \Big], \ee where
$\omega(t) = \sqrt{a + 2q\cos\,2t}$.

\subsection{Mathieu-Kapitza equation}

%==============================
\begin{figure}
\begin{subfigure}{.5\textwidth}
  \centering
  % include first image
  \includegraphics[width=5.9 cm]{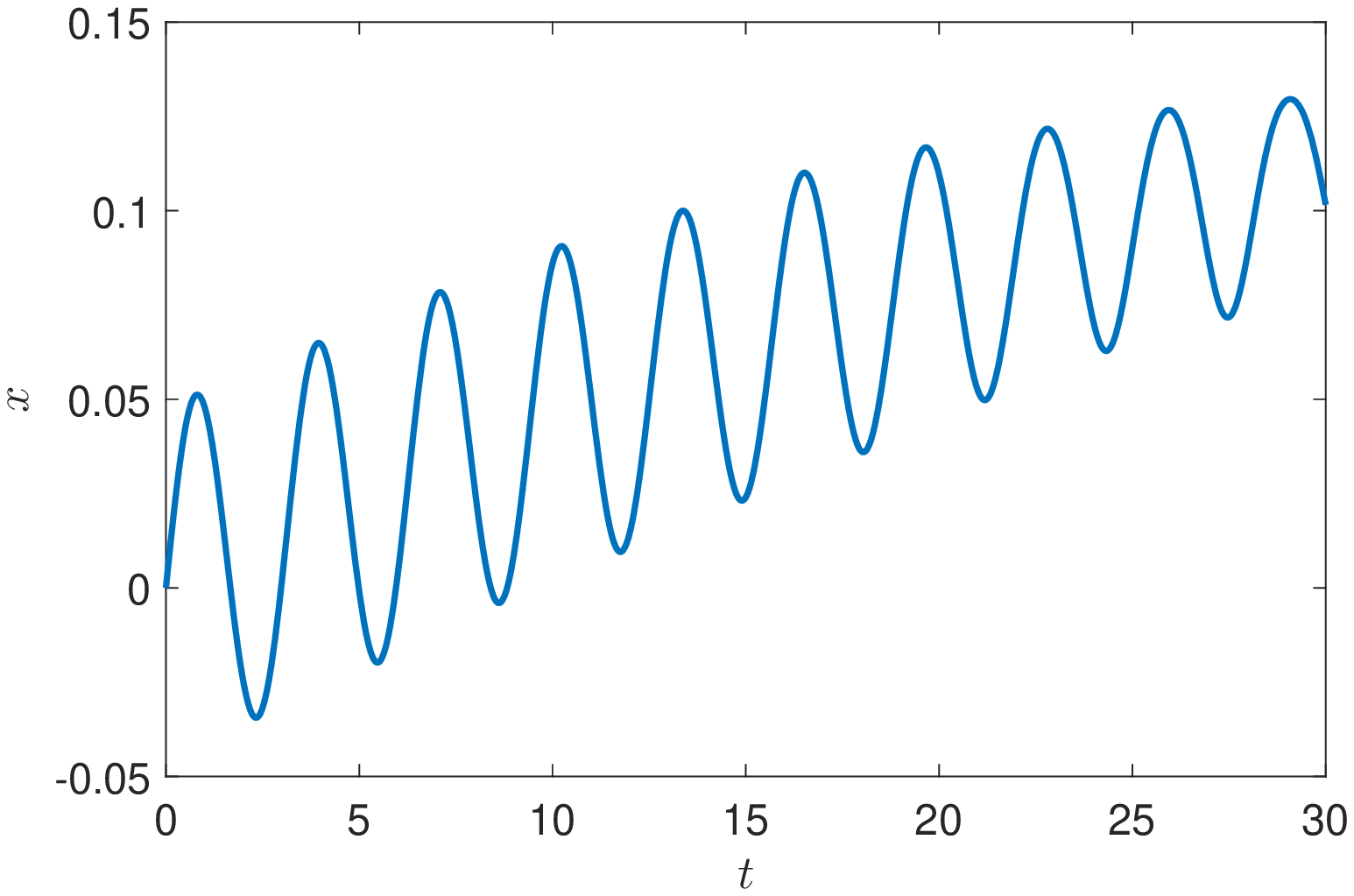}
  \caption{Plot of $x$ vs $t$.}
  \label{f3a}
\end{subfigure}
\begin{subfigure}{.5\textwidth}
  \centering
  % include second image
  \includegraphics[width=5.9 cm]{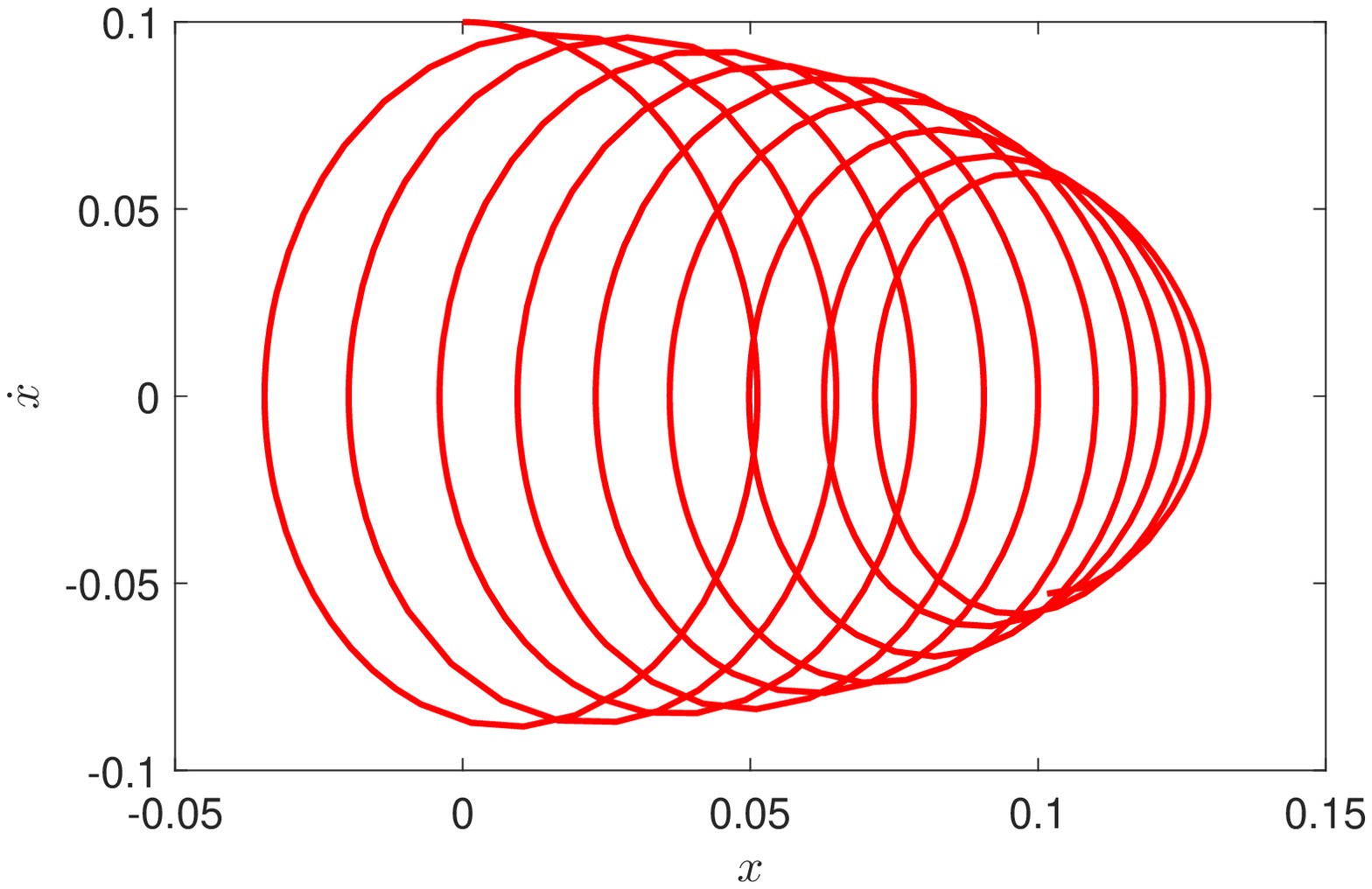}
  \caption{Plot of $\dot{x}$ vs $x$.}
  \label{f3b}
\end{subfigure}

%\newline

\begin{subfigure}{.5\textwidth}
  \centering
  % include third image
  \includegraphics[width=5.9 cm]{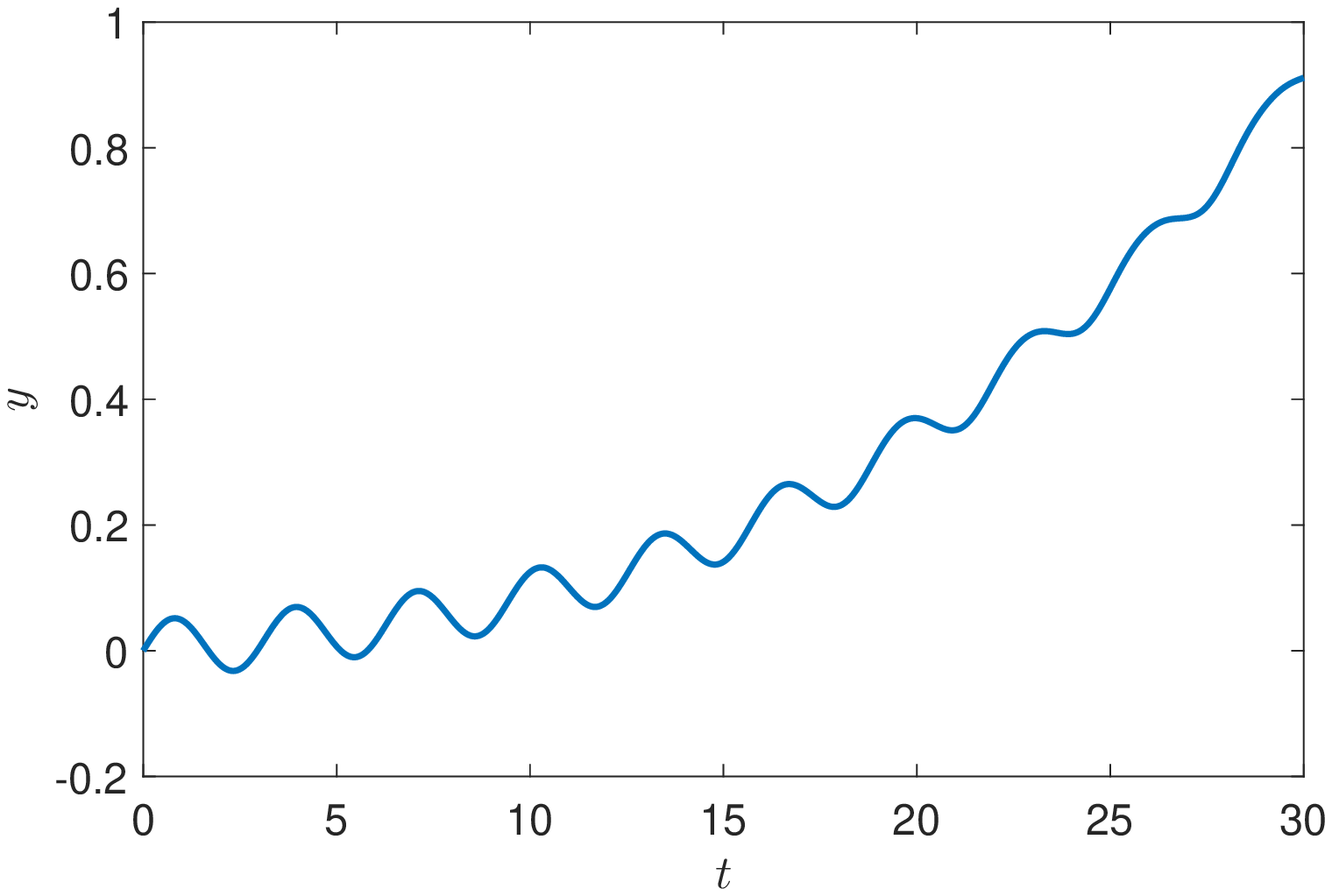}
  \caption{Plot of $y$ vs $t$.}
  \label{f3c}
\end{subfigure}
\begin{subfigure}{.5\textwidth}
  \centering
  % include fourth image
  \includegraphics[width=5.9 cm]{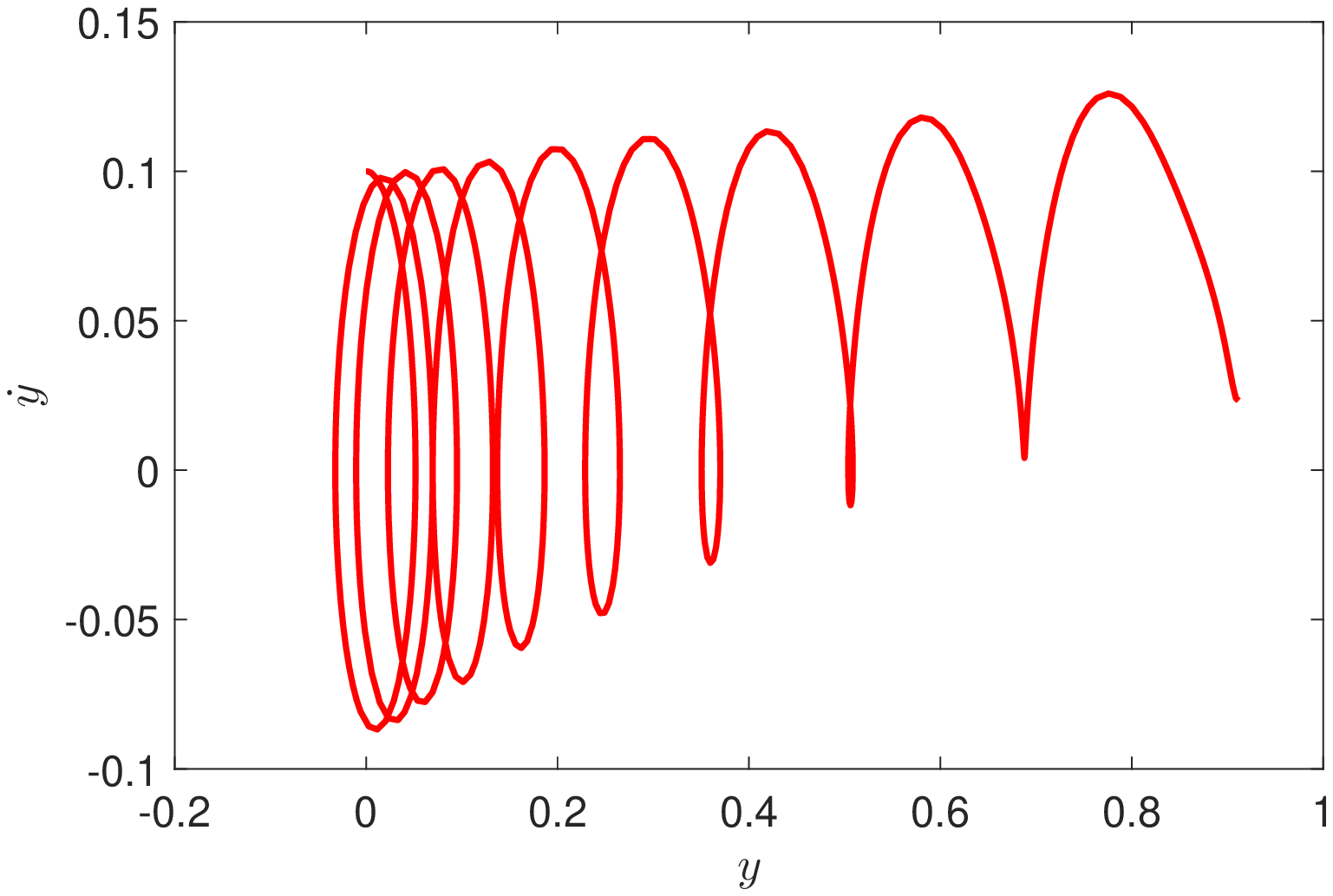}
  \caption{Plot of $\dot{y}$ vs $y$.}
  \label{f3d}
\end{subfigure}
\caption{Trajectories and phase space diagram solved from Eqs.(\ref{mk1},\ref{mk2}) with $q =0.1$, $a= 0.01$. The initial conditions are $x(0)=0$, $\dot{x}(0)=0.1$, $y(0)=0.0$, $\dot{y}(0)=0.1$.}
\label{f3}
\end{figure}
%=================================================================
%

Motivated from the above idea we can give a Mathieu modulation of the Kapitza equation
If we substitute $\omega^2(t) = (a + 2q\cos\,2t)$ in parametric Kapitza equation
we obtain
\be
\ddot{x}+ \big(\frac{4q \sin\,2t}{a + 2q\cos\,2t} \big)\dot{x} + (a + 2q
\cos 2t)^2(x + y ) = 0, \label{mk1}
\ee
\be
\ddot{y}+ \big(\frac{4q \sin\,2t}{a + 2q\cos\,2t} \big)\dot{x} - (a + 2q
\cos 2t)^2(x - y ) = 0.\label{mk2}
\ee
Clearly this pair belongs to the family of coupled linear damped Mathieu equations and the underlying potential is
the simple saddle.

\smallskip

The integral of motion of this set of equations are given by
\be
{\cal I}_1 = \frac{1}{2} \Big[\frac{\dot{x}^{2}}{a + 2q\cos\,2t} - \frac{\dot{y}^{2}}{a + 2q\cos\,2t}
+ (x^2 - y^2) + 2xy \Big],
\ee
\be
{\cal I}_2 = \frac{\dot{x}\dot{y}}{a + 2q\cos\,2t} - \frac{1}{2}(x^2 - y^2) + xy. \ee

\bigskip

Similarly we can derive the {\it nonlinear Mathieu-Kapitza equation} using the Monkey saddle. Hence using
(\ref{paraKapitza}) we obtain
\be
\ddot{x} + \big(\frac{2q \sin\,2t}{a + 2q\cos\,2t} \big)\dot{x} + (a + 2q
\cos 2t)^2 \big(x^2 - y^2 + 2xy \big) = 0, \ee
\be
\ddot{x}  + \big(\frac{2q \sin\,2t}{a + 2q\cos\,2t} \big)\dot{x} - (a + 2q
\cos 2t)^2 \big(x^2 - y^2 - 2xy \big) = 0.
\ee

In figure (\ref{f3}), we plot the space and phase parts of the dynamical variables $x$ and $y$ for the parametric Kapitza equation viz. Eqs.(\ref{mk1},\ref{mk2}). We can see that both the space parts viz. figures (\ref{f3a},\ref{f3b}) have periodicity but the amplitudes are inclining with time. The most important observation is the phase space plots viz. figures (\ref{f3c},\ref{f3d}) of the two variables $x$ and $y$. While we are having periodic cycles for the $x$ with different amplitudes, we get heteroclinic orbits for the other variable $y$.
This seems to be very interesting indeed. It is a mixed phase space with both trapping and escaping particles!

\section{Outlook}

In this paper we have re-examined integrable modulation problem
once formulated by Moser and Veselov. We have employed the algorithm given by Bartuccelli and
Gentile to construct parametric generalization of Kapitza equation of rotating shaft.
It is known that potential
of the Kapitza equation is associated to simple saddle and its $\pi/2$ degree form.
We generalized the Kapitza equation and its parametric generalization using immediate
next higher-order saddle, monkey saddle. We illustrated our construction for
$\omega^2(t) = a + 2q\cos\,2t$  and demonstrated how one can combine the Kapitza equation
with the Mathieu equation, which we coined as Mathieu-Kapitza equation. It is noteworthy that the Kapitza family
of equation yields curl forces as pointed out by Berry and Shukla. The numerical demonstration of the parametric equations shows various natures of the phase space trajectories and also opens up the novelty in particle trapping. The introduction of the parametric curl forces in different cases, as discussed, shows that the trapping can either be well and truly on the way or it can be escaped from the system. The tuning of the parameter plays a very crucial role in confining such particles and in this sense we argue that out novel findings may play a bigger role in it. In this sense we can say that this paper demonstrated a parametric generalization of the curl force problem which in turn be a very useful in explaining many of the laboratory observations regarding particle trapping. Also the analogous Airy function solutions to the modulated Mathieu equation is another novel finding in this present context. Lastly, by following the methodology proposed by Gibbons and his coworkers, we also provided a geometric description of the integrable modulated oscillator using the Eisenhart-Duval lift.

\section*{Acknowledgments}

PG is deeply indebted to express his sincere gratitude and grateful thanks to \emph{Sir Professor Michael Berry} for his enlightening discussions regarding the Airy function solutions to the Mathieu equations and also for his constant encouragements towards this work. PG \& SG are also grateful to \emph{Professors Jayanta Bhattacharjee, Michele Bartuccelli, Guido Gentile, Anindya Ghose-Choudhury, Sumanto Chanda, and Pragya Shukla} for their interests and valuable discussions.

\section*{Declarations}

PG thanks \emph{Khalifa University} for its continued support. SG thanks \emph{Diamond Harbour Women's University} for providing the necessary research environment with constant encouragement and support.

\section*{Conflicts of interest} None.

\section*{Availability of data and material (data transparency)} Not applicable.

\end{document}